\documentclass[lengthestimate,tighten,amssymb,prb,aps,twocolumn,floatfix,tightenlines,superscriptaddress]{revtex4-1}
\usepackage{graphicx,subfigure,color}
\newcommand{\ket}[1]{\left | #1 \right \rangle}

\usepackage[T1]{fontenc}
\usepackage[latin9]{inputenc}
\usepackage{longtable}
\usepackage{graphicx}
\usepackage{color}



\begin{document}

\title{A simple scheme for expanding photonic cluster states for quantum information}

\author{P. Kalasuwan$^1$, G. Mendoza$^{1,2}$, A. Laing$^1$ T. Nagata$^{3,4}$,  J. Coggins$^1$, M. Callaway$^1$ , S. Takeuchi$^{3,4}$ A. Stefanov$^5$, and J. L. O'Brien$^{*,}$}

\address{University of Bristol, H.H. Wills Physics Laboratory \& Department of Electrical and Electronic Engineering,
University of Bristol, Merchant Venturers Building, Woodland Road,
Bristol, BS8 1UB, UK.\\$^2$California Institute of Technology, Pasadena, CA
91125, USA.\\
   $^3$Research Institute for Electronic Science, Hokkaido
University, Sapporo 060-0812, Japan. \\$^4$The Institute of
Scientific and Industrial Research, Osaka University, Mihogaoka
8-1, Ibaraki, Osaka 567-0047, Japan\\$^5$Federal Office of
Metrology, METAS, Laboratory Time and Frequency,
Switzerland. }

\date{\today}

\begin{abstract}
We show how an entangled cluster state encoded in the polarization
of single photons can be straightforwardly expanded by
deterministically entangling additional qubits encoded in the path
degree of freedom of the constituent photons. This can be achieved
using a polarization--path controlled-phase gate. We
experimentally demonstrate a practical and stable realization of
this approach by using a Sagnac interferometer to entangle a path
qubit and polarization qubit on a single photon.  We demonstrate
precise control over phase of the path qubit to change the
measurement basis and experimentally demonstrate properties of
measurement-based quantum computing using a 2 photon, 3 qubit
cluster state.
\end{abstract}
\maketitle

%\emph{Introduction.---}
Quantum information science\cite{nielsen} promises both profound
insights into the fundamental workings of nature as well as new
technologies that harness uniquely quantum mechanical behavior
such as superposition and entanglement. Perhaps the most profound
aspect of both of these avenues is the prospect of a quantum
computer---a device which harnesses massive parallelism to gain
exponentially greater computational power for particular tasks. In
analogy with a conventional computer, quantum computing was
originally formulated in terms of quantum circuits consisting of
one- and two-qubit gates operating on a register of qubits which
are thereby transformed into the output state of a quantum
alogrithm\cite{nielsen}.  In 2001 a remarkable alternative was
proposed in which the computation starts with a particular
entangled state of many qubits---a cluster state---and the
computation proceeds via a sequence of single qubit measurements
from left to right that ultimately leave the rightmost column of
qubits in the answer state \cite{ra-prl-86-5188}.

Of the various physical systems being considered for quantum information science, photons are particularly attractive for their low noise properties, high speed transmission, and straightforward single qubit operations \cite{ob-sci-318-1567}; and a scheme for non-deterministic but scalable implementation of two-qubit logic gates ignited the field of all-optical quantum computing\cite{kn-nat-409-46}. In 2004 it was recognized that cluster states offered tremendous advantages for this optical approach \cite{ni-prl-93-040503,yo-prl-91-037903}: Because preparation of the cluster state can be probabilistic, non-deterministic logic gates are suitable for making it, removing much of the massive overhead associated with near-deterministic logic gates. %These, and other techniques that dispense with CNOT gates entirely \cite{br-prl-95-010501}, reduce the resources required by 3-4 orders of magnitude.

Soon after these  theoretical developments there were
groundbreaking demonstrations of small-scale algorithms operating
on four photon cluster states \cite{wa-nat-434-169,pr-nat-445-65};
cluster states of up to six photons were produced
\cite{ki-prl-95-210502,lu-nphys-3-91}; and the importance of high
fidelity was quantified \cite{to-prl-100-210501}. It has been
recognized that encoding cluster states in multiple degrees of
freedom of photons may provide advantages to computation
\cite{jo-pra-76-052326} and has been demonstrated as a promising
route to high count rates and larger cluster states
\cite{ch-prl-99-120503,va-prl-100-160502}. However, these
demonstrations have relied on a sandwich source or double pass
crystal to create the cluster state, making their production
unwieldy, and scalability an issue. Here, we propose and
demonstrate a simple scheme which enables a path encoded qubit to
be added to any photon in a polarization encoded cluster state.
This is achieved using deterministic controlled-phase (CZ) gate
between a photon's polarization and path. We use a Sagnac
interferometer architecture that provides a stable and practical
realization of this scheme and demonstrate simple
measurement-based operations on a 2 photon, 3 qubit cluster state
with high fidelity.

\begin{figure}
\includegraphics[width=\columnwidth]{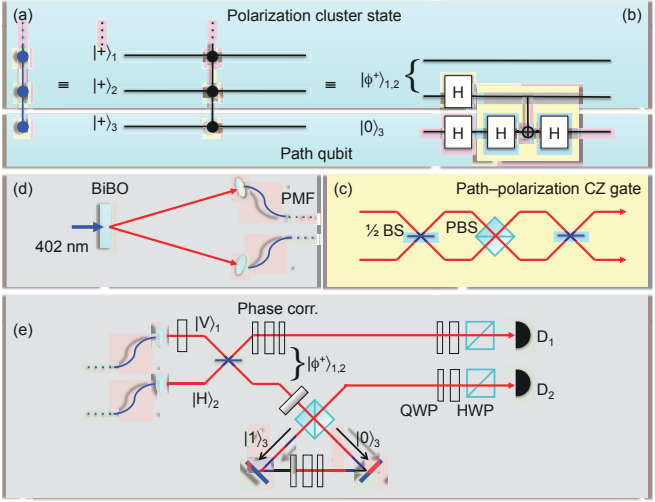}
\caption{A simple scheme for adding photon path qubits to a polarization cluster state. (a) %A cluster state is represented by a graph with nodes corresponding to qubits and bonds corresponding to entanglement links. Our goal is to add a photon path qubit to a polarization cluster state.
The linear three-qubit cluster state can be created by preparing three qubits in the $\ket{+}\equiv\ket{0}+\ket{1}$ state and implementing a two-qubit controlled-phase (CZ) gate between each. (b) The same cluster state can be realized if we start with the state $\ket{\phi^+}_{1,2}\ket{0}_{3}\equiv(\ket{00}+\ket{11})_{1,2}\ket{0}_3$, implement $\hat{H}_2\otimes\hat{H}_3$, followed by CZ$_{2,3}$. %If the maximally entangled state
(c) A controlled-NOT (CNOT) between the path and polarization of a single photon is straightforwardly implemented with a polarizing beam splitter (PBS); a CZ is realized by performing a $\hat{H}$ on the target before and after the CNOT, which for a path qubit is a 1/2 beamsplitter (BS). (d) A pair of photons were produced via Type-I spontaneous parametric downconversion in a non-linear BiBO crystal: a 60 mW 402 nm `pump' laser is shone into the BiBO; a single pump photon can spontaneously `split' into two `daughter' photons, conserving momentum and energy; degenerate pairs of photons are collected into polarization maintaining fibres (PMFs). (e) Implementation of the circuit shown in (c): %See text for details. :
the polarization entangled state $\ket{\phi^+}_{1,2}$ is realized
in post-selection by inputing a horizontal ($\ket{H}$) and
vertical ($\ket{V}$) photon into a 1/2 beamsplitter; an $\hat{H}$
on qubit 2, realized with a half waveplate (HWP), converts
$\ket{\psi^-}_{1,2}$ to the two qubit cluster state,
$(\ket{0+}+\ket{1-})_{1,2}$; the PBS Sagnac interferometer
implements a CZ between the path an polarization of photon 2 (up
to a local rotation of the path qubit).} \label{schematic}
\vspace{-0.5cm}
\end{figure}

%\emph{Approach.---}
A standard way to define a cluster state is via a graph where the
nodes represent qubits, initially prepared in the
$\ket{+}\equiv(\ket{0}+\ket{1})/\sqrt{2}$ state, and connecting
bonds indicate that an entangling controlled-phase (CZ) gate has
been implemented between the pair of qubits that they connect, as
in Fig. \ref{schematic}(a) (because these CZ gates commute, the
order in which they are performed is not important). Adding a path
encoded qubit on a photon in a polarization encoded cluster state
therefore requires a CZ gate to be implement between the
polarization of the photon and its path, which must have
previously been prepared in the $\ket{+}$ state (Fig.
\ref{schematic}(c)).

A polarizing beam splitter (PBS), that transmits horizontal and
reflects vertical polarizions of light, implements a
controlled-NOT (CNOT) gate on the polarization (control qubit) and
path (target qubit) of a single photon passing through it (Fig.
\ref{schematic}(c)). A CZ gate can be realized by implementing a
Hadamard ($\hat{H}$) gate ($\ket{0},
\ket{1}\leftrightarrow\ket{0}\pm\ket{1}$) on the target qubit
before and after a CNOT gate. For a path qubit a $\hat{H}$ can be
implemented with a non-polarizing 1/2 beamsplitter (BS). However,
preparation of the $\ket{+}$ state of the path (target) qubit
requires an additional $\hat{H}$, and $\hat{H}\hat{H}$ is the
identity operation $\hat{I}$; the $\hat{H}$ after the CNOT simply
implements a one qubit rotation, and is not included in our
demonstration. A PBS is therefore all that is required to add a
path qubit to a polarization cluster state. Measuring the path
qubit in an arbitrary basis, however, requires a phase shift
followed by BS, and so interferometric stability is required.

%\emph{Experimental scheme.---}
As a simple demonstration of this approach, we constructed the 3
qubit cluster state

\begin{equation}
\left|\Phi_{3}^{lin}\right\rangle
=\frac{1}{\sqrt{2}}\left(\left|+\right\rangle
_{1}\left|0\right\rangle _{2}\left|0\right\rangle
_{3}-\left|-\right\rangle _{1}\left|1\right\rangle
_{2}\left|1\right\rangle _{3}\right), \label{3cluster}
\end{equation}
where the first two qubits were encoded in the polarization of two photons and the third qubit was the path of the second photon. %, using the setup  shown in Fig. \ref{schematic}(e).
(Eq. \ref{3cluster} is locally equivalent to the usual 3 qubit
linear cluster state; simply with an $\hat{H}$ rotation applied to
qubit 3.) Our experimental scheme is shown schematically in Fig.
\ref{schematic}(e): Two photons prepared in the state
$\ket{1H}_1\ket{1V}_2$ converge onto a 1/2 beamsplitter,
non-deterministically creating the entangled state
$\left|\phi^+\right\rangle \equiv(\left|1H\right\rangle_{1}\left|1H\right\rangle_{2}
+\left|1V\right\rangle_{1}\left|1V\right\rangle_{2} )/{\sqrt{2}}$, where the number $1$ inside the ket brackets indicate photon number and outside subscripts 
$1$ and $2$ denote spatial paths. Photon 2 then travels through a
half-wave plate set at $22.5^{\circ}$, which implements a
$\hat{H}$ on polarization to create the two qubit cluster state. A
third qubit is added to the cluster by adding a path degree of
freedom on photon 2: Photon 2 enters the Sagnac interferometer via
a PBS cube, and forms a superposition of clockwise ($C$) and
counterclockwise ($D$) paths. The state becomes then

\begin{eqnarray}
\left|\psi\right\rangle =(\left|1H\right\rangle_{1}\left|1H\right\rangle_{C}
-\left|1H\right\rangle_{1}\left|1V\right\rangle_{D} \nonumber \\-\left|1V\right\rangle_{1}\left|1H\right\rangle_{C}
-\left|1V\right\rangle_{1}\left|1V\right\rangle_{D} )/{2}\label{eq:}
\end{eqnarray}
The relabeling $\ket{1H}_1\rightarrow\ket{1}_1$,
$\ket{1V}_1\rightarrow\ket{0}_1$,
$\ket{1H}_C\rightarrow\ket{1}_2\ket{0}_3$,
$\ket{1V}_D\rightarrow\ket{0}_2\ket{1}_3$ gives the state of Eq.
\ref{3cluster}.

The phase of the path qubit, qubit 3, can be controlled by the quarter and half waveplates (HWPs) inside the Sagnac interferometer; while the stability of this phase is provided by the Sagnac architecture  (the visibility of the Sagnac interferometer was 99.5\%). %, and so the entangling operation of the path qubit itself was performed with very high fidelity).
The angle $\alpha$ of the HWP in  the interferometer sets the
relative phase between $\ket{0}_3$ and $\ket{1}_3$ to
$e^{i4\alpha}$. The measurement basis of qubit 3 is therefore
determined by $\alpha$.

Following the principles of cluster state quantum computation, an
arbitrary qubit rotation can be performed on qubit 3 (path qubit{,
$j=3$}) by measuring qubits 1 and 2 (polarization qubits) in the
basis $B_{j}(\varphi)\equiv\{
\left|\varphi_{+}\right\rangle_{j},\left|\varphi_{-}\right\rangle
_{j}\}$ where
$\left|\varphi_{\pm}\right\rangle_{j}\equiv\frac{1}{\sqrt{2}}(\left|0\right\rangle
_{j}\pm e^{-i\varphi}\left|1\right\rangle _{j})$. The eigenvalues
$m_{j}=0$ or $m_{j}=1$ if the measurement outcome on qubit $j$ is
$\left|\varphi_{+}\right\rangle _{j}$ or
$\left|\varphi_{-}\right\rangle _{j}$, respectively. {The feed
forward information of $m_1$ selects the projection of the second
qubit: for $m_1=0$ ($m_1=1$) qubit 2 will be projected on
$\left|\varphi_{+}\right\rangle_2$($\left|\varphi_{-}\right\rangle_2)$.}
After these measurements, qubit 3 is in the state
{$\left|\psi\right\rangle_{3}=\sigma_{x}^{m_{2}}\sigma_{z}^{m_{1}}R_{x}\left(\varphi_{2}\right)R_{z}\left(\varphi_{1}\right)\left|+\right\rangle$}
Hence, the path qubit can be projected into any state (up to a
known $\sigma_x$ operation).
%In the case where the polarization of qubit1 is measured in the $B_{1}(0)$ basis, measurement of positive eigenvalues results in the spatial qubit always being found in the D path (\emph{i.e.} the $\ket{1}_3$ state).
The waveplate settings in front of the PBSs determine %the basis of measurements on qubits 1 and 2, and thereby
$\varphi_1$ and $\varphi_2$; simultaneous detection of the two
photons at detectors $D_1$ and $D_2$ ideally results in a
sinusoidal interference fringe, as a function of $\alpha$, with a
phase and amplitude that depends on $\varphi_1$ and $\varphi_2$.%on the given measurents on qubits 1 and 2.
%Since the measurement of polarization qubit 2 takes place in the C path, coincident detection of photons 1 and 2 should show a clear sinusoidal interference fringe as a function of $\alpha$.
%A measurement-based quantum operation using the cluster state of Eq.~\ref{3cluster}can be verified by measuring the polarization qubits in the $B_{j}(\pi)$ and $B_{j}(0)$ bases. Measuring positive eigenvalues will cause the spatial qubit to always be found in the D path. Since the measurement of polarization qubit 2 takes place in the C path, coincidence counts should show a clear sinusoidal interference fridge.

\begin{figure}%[t!!!]
\includegraphics[width=\columnwidth]{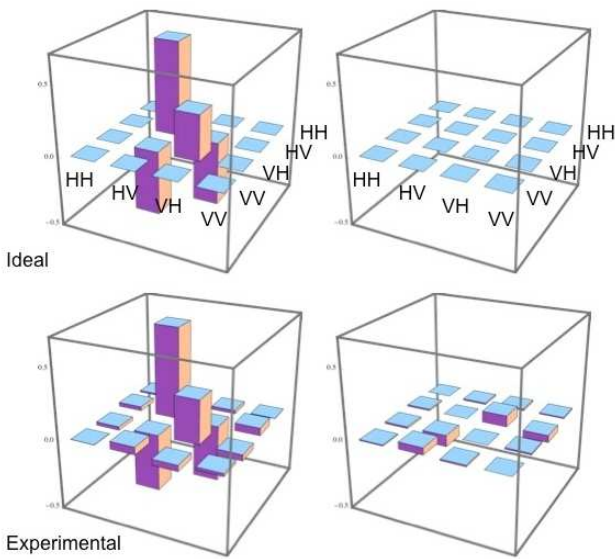}

\vspace{-0.4cm} \caption{Real (left) and imaginary (right) parts
of the experimentally measured density matrix $\rho_{exp}$
(bottom), which has a fidelity of $F= 0.929 $ with expected state
$|\psi'\rangle$ (top).} \label{dm} \vspace{-0.4cm}
\end{figure}

Figure~\ref{dm} shows the density matrix $\rho_{exp}$, obtained via quantum state tomography, of the polarization state of the two photons after the ordinary BS in Fig.~\ref{schematic}(e), before the path qubit is added. (Here the phase correction waveplates were set to produce the singlet state $\ket{\psi^-}\equiv(\ket{01}-\ket{10})/\sqrt{2}$, rather than $\ket{\phi^+}$).  It has a fidelity with the singlet state $\ket{\psi^-}\equiv(\ket{HV}-\ket{VH})/\sqrt{2}$ of $F=0.895$. A major source of this non-unit fidelity is that the BS had a reflectivity of $R=0.59$; the fidelity of $\rho_{exp}$ with the expected output state $\ket{\psi'}\equiv 0.57\ket{HV}+0.82\ket{VH}$ is $F=0.929$. The remaining imperfections predominantly arise from the non-unit visibility of quantum interference at the ordinary BS: the measured visibility for two photons of the same polarization was $V_{meas.}=0.91$, which is $V_{rel.}=0.97$, relative to the ideal visibility for a $R=0.59$ BS $V_{ideal}=0.937$. This visibility results in reduced coherences in the measured density matrix shown in Fig.~\ref{dm}. %, where the $R=0.59$ BS gives rise to the imbalance in $\ket{HV}$ and $\ket{VH}$ populations.
These imperfections  in $\rho_{exp}$  will limit the performance of cluster state operations described below. %; incorrect phases can be corrected before the interferometer.
%olinmion the performance of the cluster state operations below, and can effect the phase dependence of the projective measurements on the polarization qubits. <I don't understand this point---job>  The experimental density matrix can be improved using phase correction before the interferometer. However,  unfixable factor is the imperfection of the beam splitter.

%In practice, however, a global phase error may be present within the interferometer. This error can be offset by varying $\alpha$ in the interferometer. Furthermore, the different states of path qubit pronounced the different offset phase on the fringe. This experiment exploited these differences to distinguish states of the path qubit.

\begin{figure}[t!]
\includegraphics[width=\columnwidth]{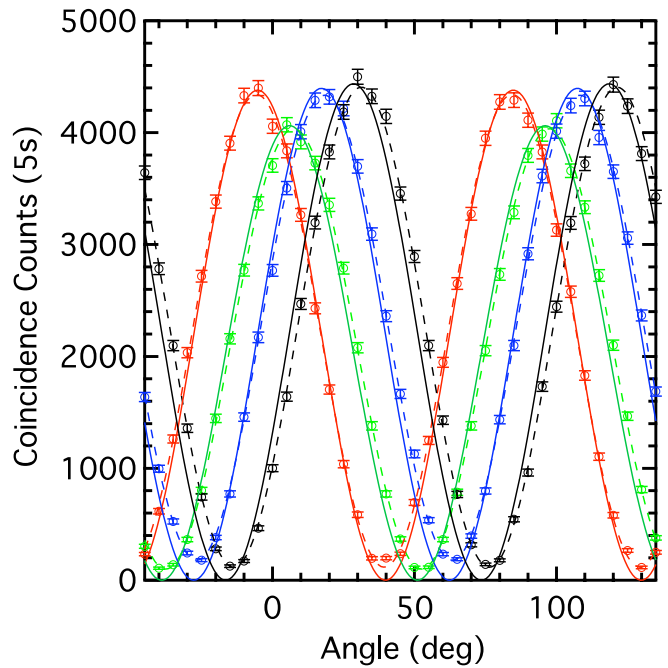}
\caption{{Path qubit rotation via polarization qubit measurement
superimposed on the theoretical curves.The fringes of coincidences
counts are pronounced as a function of $\alpha$ as described in
the text. The solid lines represent the theoretical prediction
given the reflectivity of our BS. The experimental points $\circ$ with errors are fitted by the dashed lines.}}
\label{fringes} \vspace{-0.4cm}
\end{figure}

Figure \ref{fringes} shows experimentally measured coincidence counts as a function of $\alpha$ for several different projective measurements on (polarization) qubits 1 and 2: $B_{1}(\pi/2)\otimes B_{2}(\pi/2)$ (red), $B_{1}(\pi/2)\otimes B_{2}(\pi/4)$ (green), $B_{1}(\pi/2)\otimes B_{2}(0)$ (blue) and $B_{1}(\pi/2)\otimes B_{2}(-\pi/4)$ (black)%, where $\ket{R}\equiv\ket{H}+i\ket{V}$, $\ket{L}\equiv\ket{H}-i\ket{V}$, $\ket{+}\equiv\ket{H}+\ket{V}$ and $\ket{-}\equiv\ket{H}-\ket{V}$
. {The solid lines are theoretical prediction of the fringe
expressed as
%\begin{equation}
$Y(\alpha)=Y_0(1+(1-2a^2)\cos(4\alpha+\varphi_{2})\\+2a\sqrt{1-a^2}\sin(4\alpha+\varphi_{2})\sin(\varphi_{1}))$,
%\label{equation:yy}
%\end{equation}
where $Y_0$ is the peak coincidences counts from each experiments
and $a(=0.567)$ is a constant depending on the
reflectivity$(R=0.59)$ of the BS. The relation between $R$ and $a$
is $a^2=(1-R)^2/\left((1-R)^2+R^2\right)$.} The expected high visibility fringes are observed in each case (the non-unit visibility is a result of the reduced coherences in $\rho_{exp}$), however the phase of each fringe is {offset (10's of degrees) compared to the case for a $R=0.5$ BS but is good agreement with $Y(\alpha)$}. Taking into account the $R=0.59$ BS well explains these offsets. %, while the reduced coherences of $\rho_{exp}$ explains the non-unit visibility of the fringes.
Similar fringes were measured for other projective measurements on
qubits 1 and 2: $\bigl\{ B_1(-\pi/4)$, $B_1(0), B_1(\pi/4)$,
$B_1(\pi/2)\bigl\}\otimes\bigl\{ B_2(-\pi/4)$, $B_2(0),
B_2(\pi/4)$, $B_2(\pi/2)\bigl\}$ (not shown), and again the
observed phases and visibilities were in good agreement with
predictions based on {an $R=0.59$ BS}. Observation of these
fringes confirms the correct one-qubit rotations are realized via
the measurements on the two-photon, three-qubit cluster state.

We have experimentally demonstrated a simple scheme for adding
path-encoded qubits to a polarization-encoded cluster state  and
demonstrated simple one-qubit rotations on such a hybrid
path-polarization cluster state. Similar approaches have used less
stable Mach-Zehnder interferometers \cite{pa-oe-15-17960}; while
10 qubits on 5 photons have been entangled in a similar way
\cite{gao-2008}. Photonic approaches to exploring cluster states
and measurment based quantum computations are currently the most
advanced. Further progress is limited by the number of photons,
making schemes for encoding more than one qubit per photon
appealing. The advent of high performance waveguide integrated
quantum circuits \cite{po-sci-320-646,marshall-2008} that include
ultra-stable interferometers \cite{po-sci-320-646,matthews-2008}
and precise optical phase control \cite{matthews-2008}, is a
promising architecture for this approach.  Our scheme uses entanglement of polarization and path degrees of freedom of one photons. 
This enables the addition of a path qubit to any photon in a polarization cluster
state. The path qubit is not fully connected in the
cluster, because the path qubit can be connected to the polarization qubit sharing same photon only. This is most useful at the edges of the cluster state.
With current approaches using up to six photons, adding path
qubits in this way has the potential to significantly increase the
size of cluster states, and thereby the complexity of algorithms
that can be implemented. However, there is some possibilities to entangle path qubits from different photons \cite{ra-pra-65-062324} to develop more sophisticate cluster state.

\vspace{0.4cm} \noindent We thank T. Rudolph, N. Yoran and X.-Q.
Zhou for helpful discussions. This work was supported by IARPA,
EPSRC, QIP IRC  and the Leverhulme Trust. G.M. acknowledges
support from Caltech's Summer Undergraduate Research Fellowship
(SURF).

\end{document}